# Design of a UE5-based digital twin platform

LYU Shaoqiu[1,3], Wang Muzhi[1], Zhang Sunrui[2], Wang Shengzhi[3]

（1,University of New South Wales(UNSW) ENG Sydney, Australia;
2, Royal Melbourne Institute of Technology(RMIT) SOD Melbourne, Australia;
3, Talent Science And Technology Co.,Ltd,　Shuike Shangyu Energy Technology Research Institute Co., Ltd. China Nanjing 21000）

**Abstracts:** Aiming at the current mainstream 3D scene engine learning and building cost is too high, this thesis proposes a digital twin platform design program based on Unreal Engine 5 (UE5). It aims to provide a universal platform construction design process to effectively reduce the learning cost of large-scale scene construction. Taking an actual project of a unit as an example, the overall cycle work of platform building is explained, and the digital twin and data visualization technologies and applications based on UE5 are analyzed. By summarizing the project implementation into a process approach, the standardization and operability of the process pathway is improved.

**Keywords: ue5; visualization; digital twin; engineering design**

With the development of the informationization in China, smart phones, computers, advertising screens and other visual interactive media are becoming more and more popular, and the public and enterprises have higher and higher requirements for digital informatization's applications. At the same time, with the rapid development of digital twin technology, the traditional modeling software in building large-scale scenes faces the challenge of cost and efficiency. The gaming engines have also entered the public's view with the popularity of video games, and their highly customizable/multi-platform compatible/3D visualization features make them a feasible solution for digital twin platforms. Among the gaming engines, Unity and Unreal Engine are the two most well-known ones, and Unreal has been updated to UE5 (Unreal Engine 5), which brings richer light/scene support. The paper will focus on discussing the engineering process in UE5-based digital twin platform, including scene construction, data processing and user interaction, and show its potential in improving the efficiency of digital twin realization and application.

As a game engine that focuses on building 3D-related scenes, UE5 has an advantage in supporting and rendering grand 3D scenes that is difficult to achieve on other platforms. Compared to Unity3D, UE5 has more complete 3D plug-in support and simulation rendering performance advantages, richer graphic terrain/material editing functions, and at the same time, it can use the built-in movie tools to provide better delivery of the display effect[1]; compared to Blender, UE5 can provide more interactive scenes and settings while being compatible with 3D models; and compared to the traditional html-based display, UE5 has the advantage of being easy to use (visual programming and real-time previewing) and more efficient. With UE5, developers can significantly improve the efficiency and cost-effectiveness of modeling while maintaining visual effects.

Before considering UE5 as a platform for digital twins, it is important to understand the shortcomings of UE5: as a game engine, the accuracy of its physics/fluid simulation is not comparable to that of professional simulation software; at the same time, it has certain performance requirements for development and display equipment.

## 1. Overall project planning and platform preparation

### 1.1 Project Planning Process

In general, the digital twin project can be divided into 3 major parts.

The first part, project preparation. Before starting to build the project, developers need to complete such as requirements analysis, risk assessment and management, node planning, project task segmentation and other preliminary work.

The second part, project construction. We mainly divide the project of building UE5-based digital twin platform into 4 major stages and 4 major sections. (See section 2.1, 2.2 for details).

After completing the project construction, validation and delivery, the project will enter the process of post operation and maintenance which is the third phase.

In this paper, we will focus on the construction process and method of building digital twin platform in a project based on UE5 and build a general procedure for engineering construction.

### 1.2 Data collection before project construction

UE5-based digital twin platform can visualize the effect of the project, which also means that compared with other traditional models, this platform needs to do more job in the information collection and planning work in advance.

The direction of the data collection in early stage is divided into the following five major parts, see Figure 1. the final presentation of the data collection previously is the visualization models of the project objects, spatial terrain model, database and data interface, human-computer interaction model and other management data, for example, in the construction of a water hub platform in a city in the Yangtze River Delta, through the drone patrol flights, satellite maps, manual measurements and other ways to obtain the model of the river channel and the sluice station. Of course, there will be more detailed adjustments according to different project requirements. Generally speaking, higher precision models are obtained through design drawings, while the rest correspond to lower precision models.

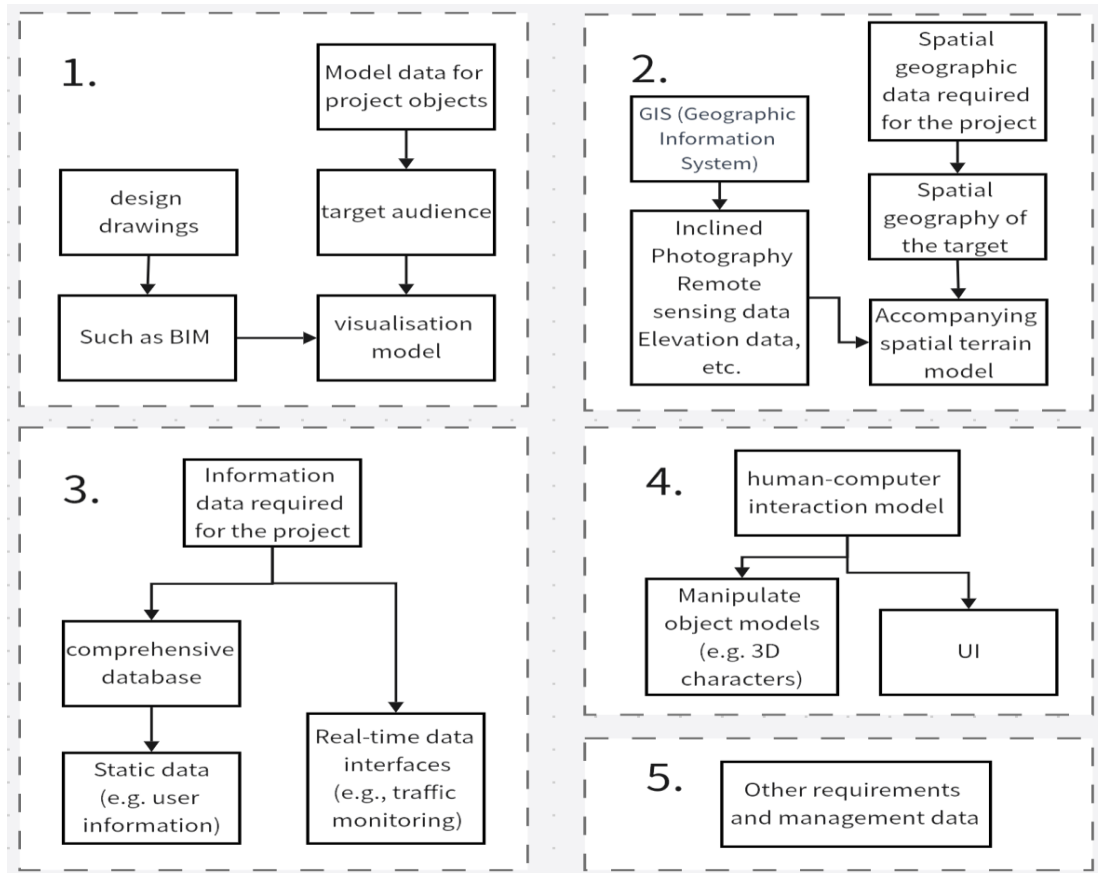

Figure 1.　5 main parts of data collection before construct the project

## 2. Design flow of platform building
### 2.1 Architectural Components of the UE5 Digital Twin Platform

According to the information visualization business requirements and UE5 platform features, in general the digital twin platform is basically divided into the following four main functions. The platform demonstrates the dynamic or static working state of the target through display roaming and scene triggering functions. Demonstrate part of the details through the simulation function or show special scenarios such as emergency fire treatment in order to visualize the scene and the treatment plan. Demonstrate the state of the project under different time, climate and light through the spatial terrain environment adjustment. Through the interactive interface UI design and trigger feedback and information guidance to reduce the difficulty of operations and enhance the user experience, the functional framework is shown in Figure 2. A water conservancy hub information broadcasting system project in the Yangtze River Delta region, has a multi-view model roaming, information monitoring, weather selection and other functions, (see section 4.3).

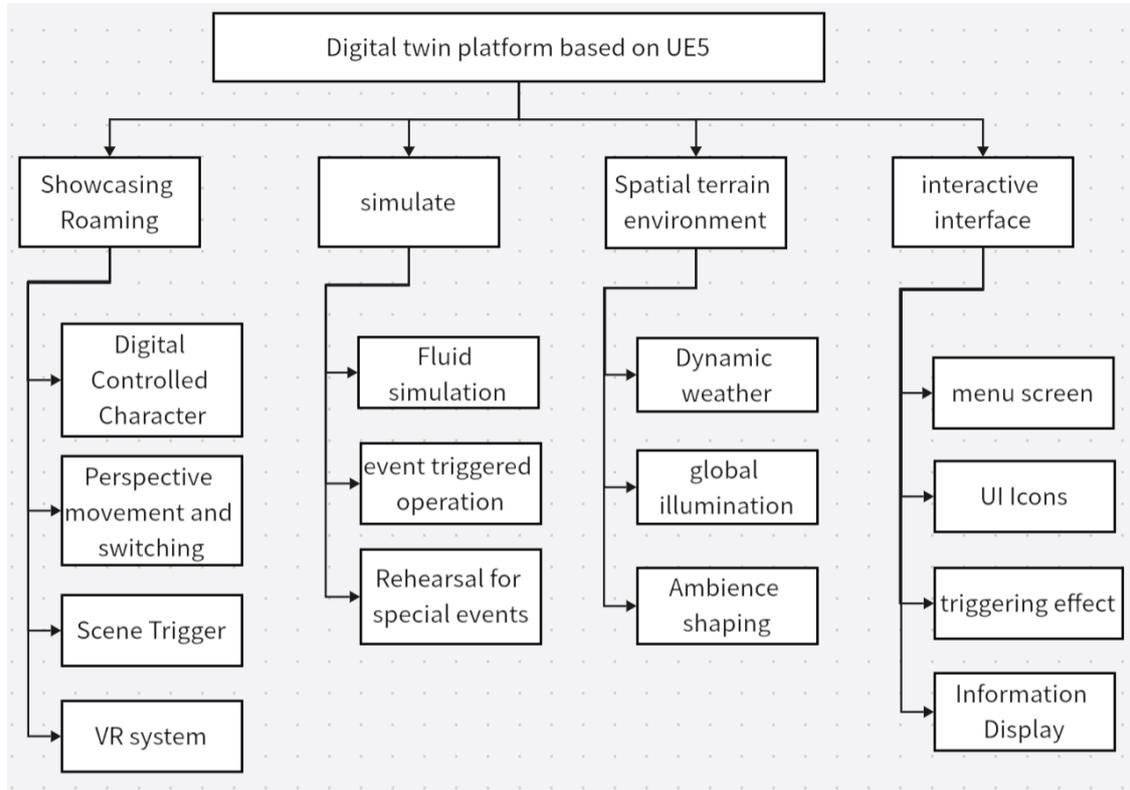

Figure 2.　Components of the UE5 digital twin platform

## 2.2 Overall platform building process

The digital twin platform intuitively displays the project's visual and other sensory effects. Its building process includes basic data collected before building, scene building in ue5 engine, functional realization of UE5-based workflow and featured functions, project testing and final project display and delivery. The specific workflow is shown in Figure 3.

1)The basic data processing part has been described in detail in the previous section, so no repeat here.
2) Scene construction based on UE5's powerful terrain rendering and large model processing capabilities, such as through the conversion of dataSmith and other plug-ins, you can easily integrate large and complex model materials and clusters. At the same time, relying on epic's huge commercial and other open-source resource model library, you can quickly build a static scene yourself.
3) Testing and final project delivery.UE5 can be deployed remotely through, for example, pixel streaming, and also supports a wide range of platforms, including PCs, game consoles, mobile devices, and VR/AR devices. This makes it more flexible to deploy and use digital twin applications on different devices and platforms.

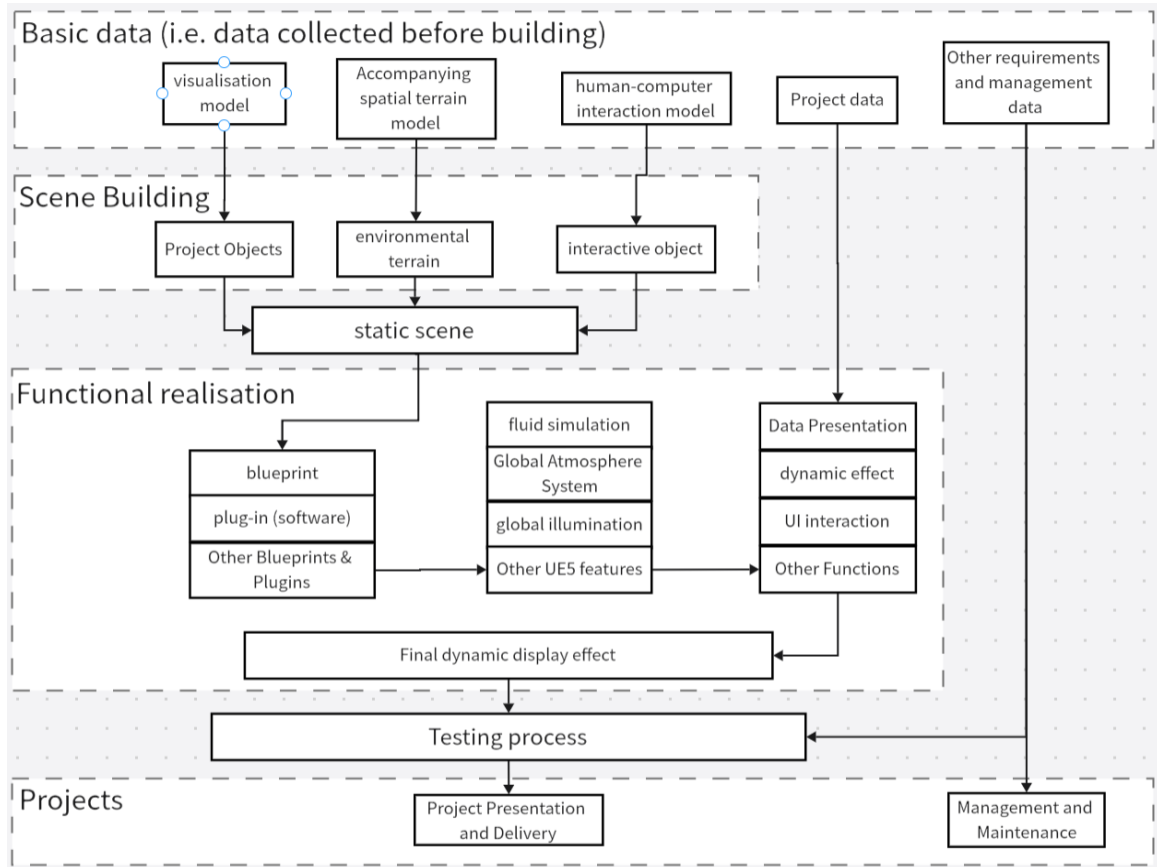

Figure 3. Technical phases of the platform's construction

## 3. Platform presentation and other project requirements and detailed technical implementations

UE5 engine has powerful features, offering many unique advantages in project usage. Below are selected unique detailed functionalities of UE5 widely used in projects.

**3.1 Visualization and interaction**

Building a digital twin desktop system through UE5's powerful built-in systems can improve user awareness and understanding of the digital twin system, thereby supporting effective management and optimization. By visualizing information and using human-computer interaction, users can intuitively understand the status, changes, and relationships of physical entities or processes in the form of graphics, charts, animations, etc., thereby better monitoring, analyzing, and understanding the operation of the system.

**3.1.1 User interface design**

In the early stage of creating a digital twin desktop system, end-users should be investigated to ensure that the interactivity of the digital twin system can meet their needs and level of understanding [2]. When completing the digital twin project through UE5, it is necessary to create the design of the system interface according to the end user habits and visual design principles [3]. The user interface design is directly affected by the quality of user experience and the ease of use of the digital twin system. In the process of designing the system interface, it is not only necessary to create the corresponding button or text through UE5's blueprint system, but also to consider the layout, structure and navigation mode of the user interface to ensure that users can clearly understand the hierarchy and organizational structure of information and quickly find the required function [4]. The overall visual interface design is equally crucial, as it should align with the style of the digital twin project, encompassing elements such as graphics, colors, and typography [5]. Furthermore, adherence to principles of visual design is essential, ensuring clarity and understanding of the interface, maintaining uniformity with subtle differences, and avoiding the occurrence of distracting elements [6]. Well-designed and visually comfortable user interface design can enhance the user's experience.

**3.1.2 User interaction design**

To enhance user interaction in using the digital twin desktop system, the powerful Blueprint system of UE5 can be leveraged to augment interactivity in digital twin projects. Designers can visually create various interactive logics, enabling users to interact with the system intuitively through actions such as clicking and dragging. For different digital twin projects, designers can set up various colliders and triggers in the scenes, and the animation system can be used to add vivid effects to the interaction and improve the user's sense of operation feedback.

For example, when designers are using UE5 to create a digital twin project for ship passage through a lock, they can add a lock start button in the digital twin desktop system. When the end user presses the lock start button, real-time status information of the lock gates will be displayed on the screen, and with the help of the animation system, real-time water level status and the ship's operational dynamics can be showcased, assisting end users in monitoring the ship's passage process in real-time.

**3.2 Data update and display**

Network protocols are crucial in digital twin engineering, where information display plays a vital role. Compared to traditional static information modes embedded within projects, UE5 offers data transmission capabilities based on network protocols, greatly expanding the project's presentation capabilities.

1) UE5 supports the HTTP protocol. Through user requests to the server, the server responds with data. The advantage of the HTTP protocol is its ability to simplify server design and implementation, making it suitable for large-scale distributed environments.

2) UE5 Blueprints support the WebSocket protocol. It establishes a persistent connection through a handshake with the server. The advantage of the WebSocket protocol is real-time bidirectional communication and lower communication latency. It can update data in the server in real-time, achieving a WYSIWYG (What You See Is What You Get) effect [7].

Overall, different methods can be chosen separately according to customer needs, or multiple methods can be mixed to meet project requirements while saving server resources. However, as of now, UE5 does not officially support WebSocket plugins, requiring third-party plugin support.

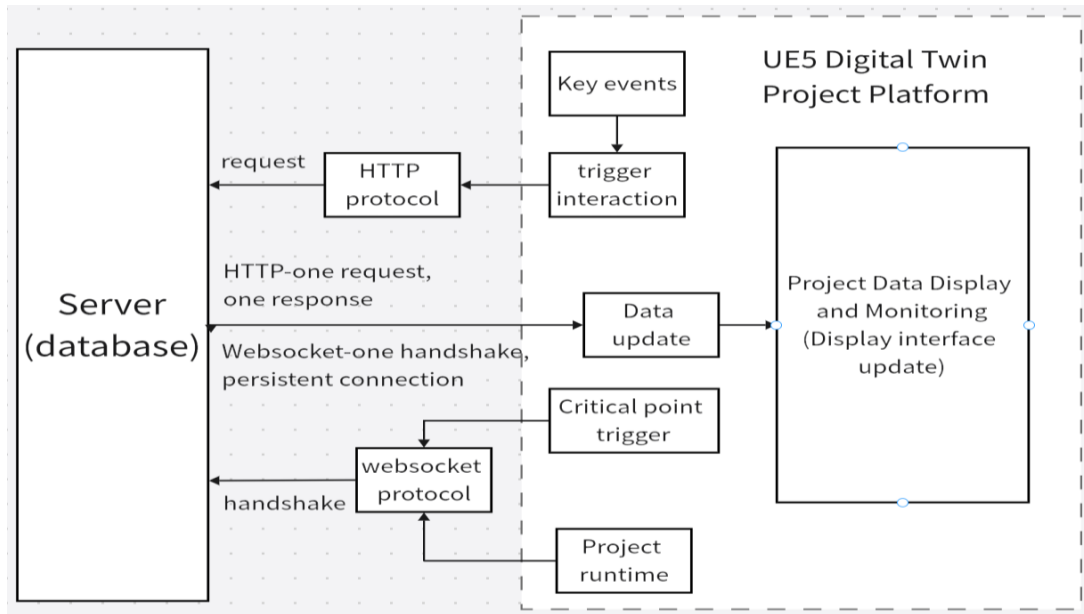

Figure 4. Technical diagram and structure of different network protocols used in UE5

## 3.3 Simulation and runtime simulation

The built-in Niagara fluid plugin in UE5 can be triggered through blueprints or key presses. This feature allows for real-time fluid physics calculations and animated displays. With the standard templates provided by the fluid plugin, various fluid models can be quickly deployed, visually showcasing scenarios involving fluids, such as dam engineering or corridor smoke systems.

For example, the water flow simulation in irrigation channels includes the display of water flow dynamics when turning the door on and off, the simulation of water flow in the pipeline, and the emergency disposal in the case of accidents. Trigger the scene demonstration through the screen button and use the fluid plug-in instead of the original discrete time axis animation effect. Compared with the conditional animation trigger, the fluid plug-in can give real-time feedback according to the behavior of the controller, and the fluid simulation greatly simplifies the cost of animation.

## 4. Partial project process demonstration

The related processes of the research results of this paper have been applied to the irrigation projects in northern Jiangsu, the Yangtze River Delta Basin, and the farmland in the middle of Jiangsu. Provide information visualization services for engineering design, plan rehearsal, facility planning, and other businesses. Through the technical improvement brought by the new engine, the technical and time costs of construction communication, software planning and development, operation and maintenance are significantly reduced, and the quality of information visualization service itself is improved.

### 4.1 Design verification of diversion channel in the project (Irrigation district project)

The UE5 platform engine supports fluid demonstration, and the Niagara fluid plug-in supports fluid precision programming, which can be customized to meet the needs of the project. The water diversion dynamic effect of the diversion tank can be intuitively displayed in the design phase.

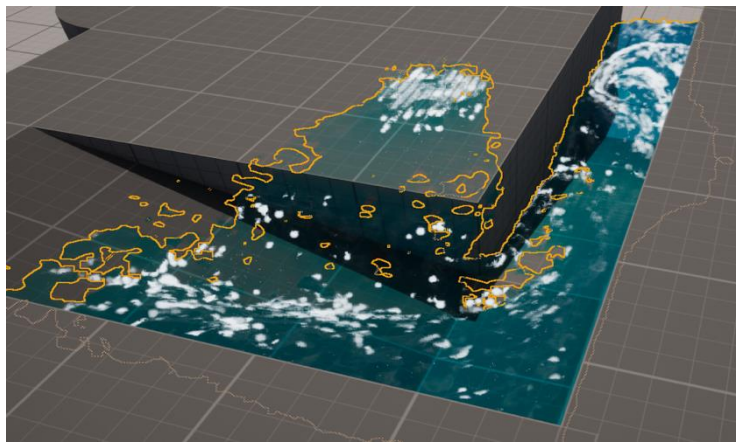

Figure 5. Diversion channel verification using UE5 Niagara plug-in

## 4.2 Character and vision design

The unique character asset editor of the digital twin platform based on UE5 greatly simplifies the construction process of characters and first-person perspectives. The ability to freely trigger and switch between multiple perspectives significantly enhances the freedom of controlling characters and immerses users in the operation.

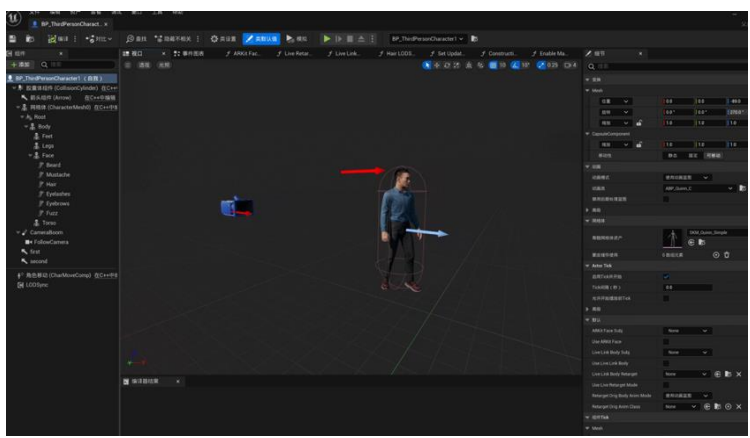

Figure 6.  Edit using the UE5 Character Editor

## 4.3 Scene interaction and project presentation

The interaction triggering functionality of the UE5 engine can be based on UE5's blueprint programming feature. The characteristics of low code and the engine can greatly reduce the threshold of work. By triggering animations, engineering interactivity and user immersion can be greatly enhanced.

Overall, the project has achieved the consolidation of multiple platform perspectives into the same interface, as seen in Figure 7. Meanwhile, for detailed project content, such as pump stations, as shown in Figure 8, the platform has implemented functionalities including operation monitoring, sluice station monitoring, water level feedback, water quality testing, and more. This has enhanced the perception capabilities of staff towards engineering projects while simultaneously reducing the patrol costs for maintenance personnel.

In addition, apart from directly editing frontend page visuals in UE5 to stream the project, the scene interaction page can also serve as a purely display frontend page, where model perspectives can be streamed over the network using UE5's pixel streaming.

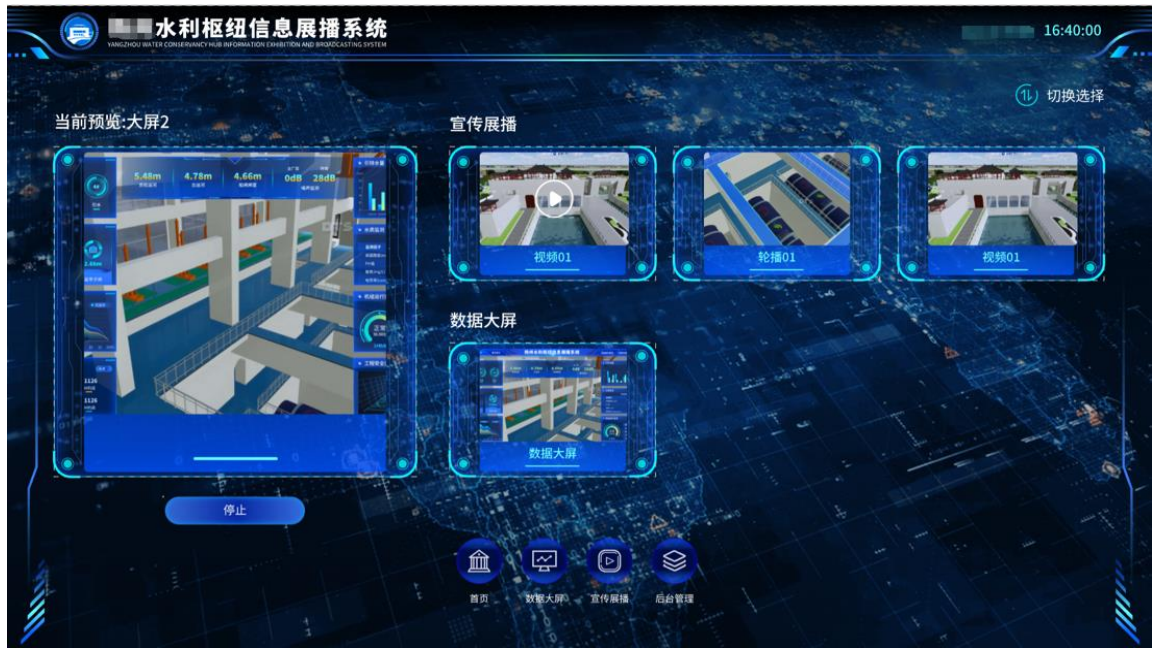

Figure 7.　Display interface of a water conservancy hub in the project

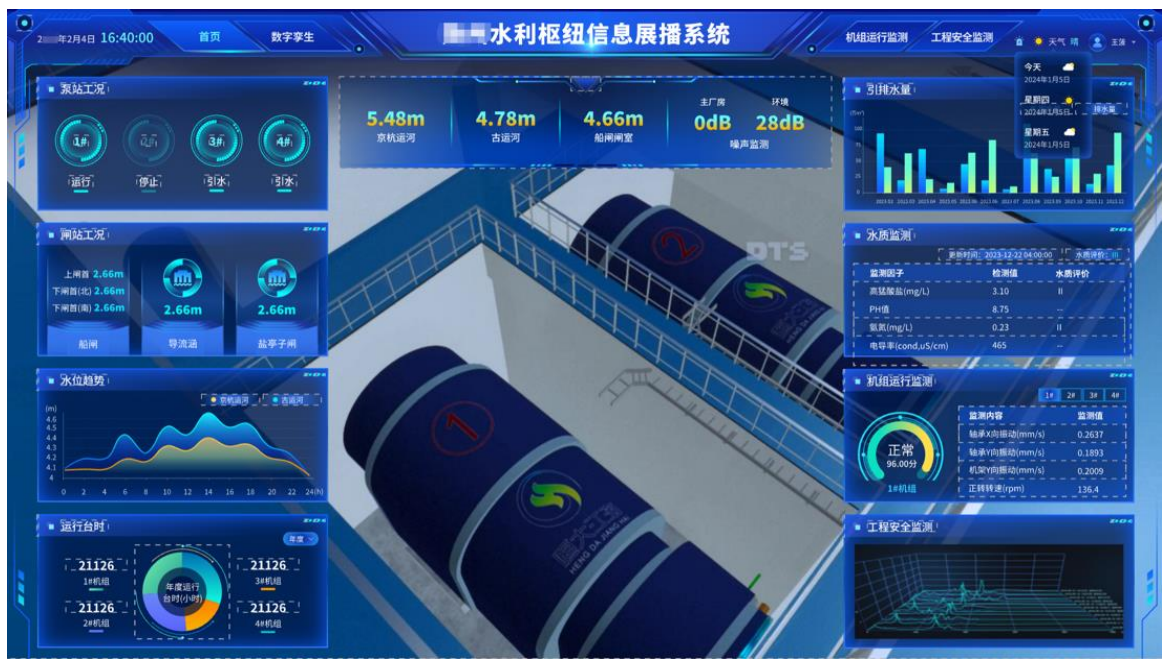

Figure 8.　Pumping station display interface in the project

## 5.Conclusions

This paper discusses the feasibility of a new digital twin platform based on the UE5 engine, and analyzes and demonstrates the relevant application scenarios and presentation forms, etc. The UE5 platform can greatly shorten the development process and development difficulty while meeting the presentation effect as much as possible and deliver the finished products in a more efficient and convenient presentation form, so it can be used as the main develop platform for enterprises to carry out digital display and simulation in the future.

This paper proposes a development process for UE5 engine for digital twin projects, which needs to carry out the application development after the complete preparation of the data collection and modeling, but the use of UE5 blueprint function can quickly complete the scene interaction/data feedback/visual presentation, and at the same time use a variety of plug-ins can simplify some of the complex functions.

UE5 supports the customization of a variety of functions and has a wide range of community open-source software and plug-in support. In the future, customization and development of features can be achieved based on open-source plugins according to specific needs, facilitating a transition from application to creation. Using the characteristics of UE5 engine for multi-terminal support, in the future a variety of different platforms (PC/VR/mobile/web) applications development can be achieved at the same time, saving the development budget and shorten the delivery time, improve the operational efficiency of enterprises.